\documentclass[11pt]{article}

\usepackage[T1]{fontenc}
\usepackage[utf8]{inputenc}
\usepackage[margin=1in]{geometry}
\usepackage{graphicx}
\usepackage{bm}
\usepackage{mathtools}
\usepackage{amssymb}
\usepackage[hidelinks]{hyperref}

\graphicspath{{./}}
\title{From Heat Stress to Perception: Interpretable Data-Driven Models of Human Thermal Sensation}

\author{%
Abed Hammoud\textsuperscript{1},
Xinjie Huang\textsuperscript{1},
Qinqin Kong\textsuperscript{2},\\
Marialena Nikolopoulou\textsuperscript{3}, and
Elie Bou-Zeid\textsuperscript{1,*}\\[0.75em]
\small \textsuperscript{1}Department of Civil and Environmental Engineering, Princeton University, Princeton, NJ, USA\\
\small \textsuperscript{2}Woods Institute for the Environment, Stanford University, Stanford, CA, USA\\
\small \textsuperscript{3}Kent School of Architecture and Planning, University of Kent, Canterbury, UK\\[0.5em]
\small \textsuperscript{*}Corresponding author: \href{mailto:ebouzeid@princeton.edu}{ebouzeid@princeton.edu}
}
\date{}

\begin{document}
\maketitle

\begin{abstract}
Heat stress indices are designed to quantify physiological thermal stress, but their relevance for inferring the thermal perception of individuals remains unclear. 
In this study, we show that thermal stress and thermal sensation often diverge, as evidenced by distinct global sensitivity patterns with respect to environmental drivers. 
Using thermal sensation votes survey data, we demonstrate that the dominant sensitivities of stress-based metrics do not align with those governing reported human thermal sensation.
Given the multitude of globally-applicable thermal stress indices and the lack of comparable general thermal sensation metrics, we develop two complementary data-driven modeling frameworks for thermal sensation. 
First, we construct polynomial chaos expansion (PCE) surrogates to represent thermal sensation as a function of meteorological variables, enabling efficient variance-based sensitivity analysis and explicit identification of influential inputs and interactions. 
Second, we develop multilayer perceptron (MLP) classifiers that capture the nonlinear and subjective nature of thermal perception, while achieving high predictive accuracy. 
The PCE models provide physically interpretable sensitivities that can explain the drivers of thermal sensation, while the MLPs offer flexible predictive capability suited to complex, heterogeneous environments.
We apply both modeling approaches at city- and continent-scales, revealing systematic differences in sensitivity structure and performance across climates. 
In particular, we find that the sensitivity of TSV-based models to the variability of meteorological conditions across geoclimatic zone encodes distinct dependencies on temperature, radiation, humidity, and wind that vary geographically, and are generally different from those of heat stress indices. 
These results suggest that stress-based indices may not fully capture the determinants of reported human thermal sensation, highlighting the need for sensation-focused predictive models and for a better understanding of the relationship between physiological stress metrics and subjective perception. 
Practically, our findings imply that heat reduction measures most suitable for mitigating health impacts under extreme heat may be different from those that can effectively make cities more comfortable and livable. The former would need to reduce the drivers of thermal stress, while the latter need to curb the distinct determinants of thermal discomfort.
\end{abstract}

\section{Introduction}
\label{sec:intro}

Heat extremes are intensifying in a warming climate, with strong implications for public health, labor productivity, energy demand and the design and operation of the built environment \cite{matthews_mortality_2025}.
The Intergovernmental Panel on Climate Change (IPCC) Sixth Assessment Report concludes with virtual certainty that hot extremes, including heatwaves, have become more frequent and intense across most land regions since the 1950s and that anthropogenic climate change is the main driver \cite{ipcc_ar6_2021}. 
These trends have accelerated efforts to define and map ``heat hazard'' using gridded meteorological products and derived indices at regional to global scales \cite{Hersbach2020, Raymond2020, Vecellio2023, Kong2025}.

A diverse family of heat indices is routinely used to translate meteorological conditions into stress metrics intended to represent ``how hot it feels''. 
Classic examples include the heat index (temperature-humidity based) \cite{steadman_assessment_1979}, wet-bulb globe temperature (combining radiative and evaporative constraints) \cite{yaglou_control_1957,kong_explicit_2022}, Universal Thermal Climate Index (UTCI) \cite{brode_deriving_2012}, and the Physiological Equivalent Temperature (PET) \cite{hoppe_physiological_1999,walther_pet_2018}. 
These indices are widely used in occupational guidance, health studies, and climate research, increasingly supported by gridded products derived from reanalyses and climate-model output \cite{DiNapoli2020, Jian2024}. 
However, the proliferation of indices reflects a fundamental challenge: heat indices are designed for different use cases, embed different physical assumptions, and can disagree substantially about which meteorological drivers matter most under a given set of conditions \cite{terjung_thermal_2015}. 
For example, commonly used indices can yield conflicting conclusions about the effect of moisture on heat stress, implying that the inferred “importance” of humidity versus temperature depends on the chosen index \cite{sherwood_how_2018,Simpson2023}.
Multi-index datasets further underline that no single index dominates across all use cases, and that index behavior can vary geographically because the controlling meteorological factors and their interactions differ by climate regime.

For many Earth system and built environment applications, the scientifically-relevant target is not only physiological heat strain but also perceived heat, since perception influences behavior and adaptive responses, as well as cooling energy demand \cite{HE2026107668}. 
Reported thermal sensation is a perceptual outcome shaped by physiological signals, prior exposure, context, and expectations, and therefore does not necessarily coincide with metrics designed to approximate thermophysiological strain.
Perceptual responses often diverge from steady-state model predictions, or heat stress indices, due to thermal history and contextual adaptation—factors that are difficult to encode within a single universal stress metric \cite{Vellei2023}.
In these contexts, sensation-based metrics derived from thermal sensation votes (TSV) provide a direct empirical representation of perceived conditions, rather than an inferred proxy based on physiological modeling \cite{Broday2019, Zheng2025}.

A practical challenge is that TSV models are typically developed based on data collected in a certain city, reflecting its unique geoclimate and the adaptation of its population. 
To the extent of our knowledge, TSV models are only available for fewer than 20 cities, mostly in more developed regions where targeted surveys exist \cite{GOLASI2018, Huang2026}.
This motivates statistical and machine-learning models that map environmental and personal variables to sensation categories or probabilities \cite{Boudreault2025}. 
Classification approaches have long been used to relate climate variables to discrete outputs in climate applications. 
More recently, simplified approaches using gradient-boosted trees have demonstrated capability in predicting thermal comfort using data from wearable technology \cite{Kim2024}. 
However, such machine learning models often trade interpretability for accuracy, making it difficult to extract actionable guidance about which inputs, or combinations thereof, govern perceived heat across scales.

Variance-based global sensitivity analysis (GSA) provides a principled framework for quantifying how uncertainty or variability in model inputs contributes to variability in model outputs, including the role of interactions. 
In practice, variance-based GSA is commonly quantified through Sobol sensitivity indices, which decompose output variance into first-order and higher-order interaction contributions.
Computing these sensitivity indices directly from complex nonlinear models can be computationally expensive, particularly when the goal is to evaluate sensitivities across large spatial domains or under multiple climate regimes. 
Polynomial chaos expansions (PCEs) offer an efficient surrogate modeling strategy: by representing a model response as an expansion in orthogonal polynomials of the inputs, PCE enables analytic computation of Sobol sensitivity indices from the expansion coefficients. 
This property is especially attractive when the scientific objective is not only prediction, but also the identification of dominant drivers and key interactions that can inform subsequent model construction and metric design.

In this study, we test the hypothesis that physiological heat stress metrics and reported thermal sensation are governed by different meteorological sensitivity structures.
Although both depend on temperature, humidity, wind, and radiation, we expect the relative importance of these inputs, and especially their interactions, to differ between stress- and perception-oriented constructs and to vary across climates.
This hypothesis motivates the following research questions:
\begin{enumerate}
\item Do the Sobol sensitivity patterns of standard heat stress indices align with those of TSV-based thermal sensation, or do they identify different controlling variables and interactions?
\item Which first-order effects and higher-order interactions dominate the variability of thermal sensation, and how do these vary across regional and continental scales?
\item Can data-driven probabilistic models provide both interpretable diagnostics of the drivers of thermal sensation and accurate TSV predictions suitable for mapping perceived heat?
\end{enumerate}
To answer these questions, we develop two complementary modeling frameworks.
The first uses polynomial chaos expansion (PCE) surrogates to compute variance-based sensitivities and diagnose the drivers and interactions that govern thermal sensation.
The second uses multilayer perceptron (MLP) classifiers to estimate probabilities of thermal sensation categories from the same meteorological inputs.
We apply these frameworks at both regional and continent scales to evaluate how sensitivity structure and predictive performance vary across climates.


\section{Data and Methods}
\label{sec:methods}

\subsection{Heat Metrics}
\label{ssec:heatMetrics}

In this study, we examine both TSV, which describes subjective human thermal sensation, and thermal stress indices, which characterize physical human thermal dynamics. 
TSV data are typically obtained by conducting surveys through questionnaires and recording people’s perceptions as discrete numerical votes. 
TSV quantifies people’s subjective perceptions on a Likert scale \cite{Likert1932Attitudes}, for example, with a 5-point scale TSV, people recorded their perception on the thermal environment with a numerical vote from $-2$ to $+2$, with unit increments, during field surveys. 
The sign means that people feel hot (+), neutral (0), or cold (-), and the magnitude reports the level of discomfort, where the smaller the magnitude, the more neutral/comfortable a person feels (e.g., +2 means very hot, and +1 means mildly hot). 
Field surveys are usually combined with in-situ meteorological measurements of temperature, humidity, wind speed, and/or radiation. 
By performing linear regression to fit a model to the collected TSV data and simultaneous meteorological measurements, TSV equations (TSV = C$_T$ T + C$_Q$ Q + C$_W$ W + C$_R$ R + C) have been developed in some regions with in-situ TSV survey data, where $C_i$ is the coefficient of variable $i$, describing the impact of a meteorological variable $i$ on human thermal perception. 
The meteorological variable $i$ can be T (air temperature, $^{\circ}$C), Q (air humidity; e.g., relative humidity in this study, \%), W (wind speed, m/s), or R (mean radiant temperature, $^{\circ}$C), and C is the fitting intercept. 
These TSV equations are usually developed for, and only apply to, certain cities since they reflect the local geoclimate and the adaptation of the local population. 
The coefficients can thus vary a lot across cities/seasons, and the underlying dynamics are difficult to explain since they involve physiological and psychological factors that vary significantly across populations and individuals. 
However, at the population level, it should be possible to develop universal TSV models if the geoclimate and population characteristics are reflected in the model coefficients.
There is an early attempt to build a globally-applicable TSV equation by considering latitude, mean annual temperature, mean temperatures of the hottest and coldest months on top of four main meteorological variables mentioned before \cite{GOLASI2018}. 
This global outdoor comfort index (GOCI) will be discussed and compared to our new approach later. 

In this study, we also selected three thermal stress indices that include sophisticated characterization of human thermal balance: standard effective temperature (SET) \cite{gonzalez_experimental_1974}, universal thermal climate index (UTCI) \cite{brode_deriving_2012}, and physiological equivalent temperature (PET) \cite{hoppe_physiological_1999,walther_pet_2018}. 
These indices characterize human-environment thermodynamic exchanges and incorporate human thermoregulation. 
Some of these indicators not only reflect the impacts of meteorological factors on human energy balance, but also represent individual factors such as metabolic activity and clothing conditions. 
Due to data availability and to ensure comparability, we made the following assumptions for two models (PET and SET) that need specified human parameters: we use an average human setting with a body height of 1.7m and a body weight of 60 kg, and two typical clothing conditions: 0.3clo in summer and 1.8clo in winter \cite{Lee2013ClothingCLIMA}. 
We prescribed the metabolic rate based on assumed age (30-year-old) and gender (female), but the sensitivity analysis to meteorological factors should not be affected significantly by this choice of baseline individual.
PET and SET were computed using pythermalcomfort \cite{TARTARINI2020100578} while UTCI was computed using Thermofeel \cite{BRIMICOMBE2022101005}.

\subsection{Meteorological and Thermal Sensation Data}
\label{ssec:metData}

We collected TSV survey data and simultaneous meteorological data from the Rediscovering the Urban Realm and Open Spaces (RUROS) project \cite{Nikolopoulou2004RUROS}. 
We also extracted from the RUROS dataset four meteorological variables that are most critical and commonly used for thermal metrics: 2-m air temperature (T), 2-m relative humidity (Q) and 10-m wind speed (W), and surface-level mean radiant temperature (R). 
The RUROS database is the most comprehensive TSV database with a large sample size (9,028) across four seasons, and both daytime and nighttime coverage. 
However, its geoclimatic coverage is limited to five European countries (Germany, Greece, Switzerland, Italy, and the UK).
Its range of meteorological conditions is, nevertheless, sufficiently broad to provide a valid foundation for our TSV model development and for testing the sensitivities of TSV and heat stress indices.  

\subsection{Polynomial Chaos Expansions}
\label{ssec:PCEs}

Polynomial chaos expansions provide a spectral representation of model responses under input uncertainty and have become a cornerstone in uncertainty quantification for high-dimensional physical and data-driven systems \cite{wiener1938homogeneous, xiu2002wiener, lemaitre2010spectral}. 
PCE expresses a model output as a series of orthogonal polynomials defined over the probability space of the uncertain or variable inputs, $\bm{\xi}$.
The dependence of a generic quantity of interest (QoI), $Y$, on the inputs is approximated by a truncated series of the following form:
\begin{equation}
    Y\left( \bm{\xi} \right) \approx \sum_{k = 0}^{P} a_k \Psi_k \left( \bm{\xi} \right),
    \label{eq:PCE}
\end{equation}
where $\Psi_k$ denotes the orthogonal basis function defined \textit{a priori} over the probability space associated with $\bm{\xi}$ and determined from the probability measures of these inputs. 
The coefficients $a_i$ represent the expansion coefficients to be determined from the data (e.g.\ regression to obtain the most accurate fit), and $P+1$ indicates the size of the approximation basis.

For the heat metrics problem, the input variables (T, R, W, Q) are parameterized as $4$ independent canonical random variables, normalized to be uniformly distributed over $\left[-1, 1 \right]^d$.
Note that the variable inputs admit the same physical range as the one provided by the data (ERA5 and RUROS), and a Rosenblatt transformation was used to map the data from the physical range and respective distribution to the normalized one that is uniformly distributed \cite{Rosenblatt1952, Feinberg2015}. 

We rely on the standard total order truncation technique, such that the truncated basis size is:

\begin{equation*}
    P + 1 = \frac{\left( d + p \right) !}{d!\,p!},
\end{equation*}
where $p$ denotes the maximal polynomial degree of the truncated expansion.
The PCE models were constructed using a non-intrusive method to determine the PC coefficients \cite{Eldred2009, Bruno2021}.
Non-intrusive approaches are attractive because PC coefficients are determined using the outputs from deterministic forward runs that are performed for a collection of representative realizations of the uncertain inputs. 
Previous studies have provided an extensive overview of different non-intrusive methods for generating PC surrogates \cite{Najm2009a, Hadigol2018, HANTOUCHE2020355, Bruno2021}.

In the present study, the PC coefficients were determined using a regularized regression approach using the least absolute shrinkage and selection operator (LASSO) algorithm \cite{lasso1996, Roth2004, hammoud2023}.
Surrogates were constructed using this regularized regression approach since it is robust to noisy estimates, enabling a more reliable construction of PC surrogates.
The coefficients are then obtained by minimizing the sum of the squared residuals between PC predictions and observed variables, and $\ell_1$ regularization is included to avoid overfitting.
The loss to be minimized is as follows:
\begin{equation}
    \mathcal{L}(a_k) =  \left( \frac{1}{2N_o} \sum_{i = 1}^{N_o} \left( y_i - \sum_{k = 0}^{P} a_k \Psi_k\left(\bm{\xi}^{(i)}\right) \right)^2 + \mu \sum_{k = 1}^{P} |a_k| \right),
\end{equation}
where $N_o$ represents the number of observations; $y_i$ indicates the heat metric observation $i$; $\bm{\xi}^{(i)}$ denotes the i$^{th}$ input vector; and $\mu$ represents a positive regularization constant that is determined by empirical tuning. 
The optimal value of $\mu$ yields the smallest mean squared error over the validation dataset. 

\subsection{Sobol Sensitivity Indices}
\label{ssec:sobol}

Global sensitivity analysis aims to quantify how variability in model inputs propagates to variability in QoIs, i.e. outputs. 
Among the most widely used variance-based methods are Sobol sensitivity indices, which decompose the total output variance into contributions associated with individual input parameters and their interactions. 

Let $Y(\boldsymbol{\xi})$ denote a scalar QoI depending on a $d$-dimensional vector of independent random variables $\boldsymbol{\xi} = (\xi_1,\dots,\xi_d)$, with index set $D \equiv \{1,\dots,d\}$. 
Following \cite{sobol_sensitivity_1993} and \cite{Crestaux2009}, the Sobol sensitivity index associated with a subset of variables $\boldsymbol{i} \subset D$ is defined as
\begin{equation}
    S_{\boldsymbol{i}} \equiv 
    \frac{\mathbb{V}_{\boldsymbol{i}}(Y)}
         {\mathbb{V}(Y)}
    =
    \frac{\mathbb{V}\!\left(
        \mathbb{E}\left[Y \mid \boldsymbol{\xi}_{\boldsymbol{i}}\right]
    \right)}
    {\mathbb{V}(Y)},
\end{equation}
where $\mathbb{E}[\cdot]$ and $\mathbb{V}[\cdot]$ denote the expectation and variance operators, respectively. The index $S_{\boldsymbol{i}}$ measures the fraction of the total output variance explained by the subset of variables $\boldsymbol{\xi}_{\boldsymbol{i}}$ acting alone, i.e., excluding interactions with the remaining inputs.

The Sobol sensitivity indices could be estimated efficiently using the PC expansion by exploiting the orthogonality of the PC basis~\cite{sobol_sensitivity_1993, HOMMA19961, SOBOL2001271, SUDRET2008964, Crestaux2009, Alexanderian2012}. 
The Sobol indices are given by:

\begin{equation} 
    S_{\bm{i}} \approx \frac{\sum_{k\in \mathcal{C}_{\bm{i}}} a_k^2 || \Psi_k ||_2^2  }{\sum_{k \geq 1} a_k^2 || \Psi_k ||_2^2  },
\end{equation}
where $\mathcal{C}_{\bm{i}} = \{ k \, | \, \Psi_k \text{ has degree } 0 \text{ in all } \xi_j \text{ with } j \notin \bm{i} \}$, where $S_{\bm{i}}$ accounts for the direct contributions of the uncertain parameters $\bm\xi_{\bm{i}}$.

\subsection{Classification}
\label{ssec:logReg}

Classification is a statistical tool used to estimate the probability of a categorical outcome, such as the dichotomy of whether an individual feels hot or not. 
Standard classification, and in particular logistic regression, is formulated as a generalized linear model that estimates the class probabilities directly from a linear combination of input variables using the softmax transformation \cite{Bishop_2006}. 
This formulation can be further extended to nonlinear classification models, where neural network architectures are incorporated, and learned latent features enhance the classification. 

Consider an input vector $\bm{x}\in \mathbb{R}^d$ that is propagated through a multilayer perceptron model $\bm{f}_\theta: \mathbb{R}^d \rightarrow \mathbb{R}^K$, where $d$ is the input dimension, $K$ the number of classes, and $f_\theta$ is the predictive model parameterized by $\theta$.
The model ($f_\theta$) outputs a raw score vector $\bm{z}=\left(z_1,..., z_K\right)$ for each input sample. 
To interpret this score, we apply the softmax function to recast these scores to class probabilities ($\text{pr}\left(y=k|\bm{x}, \theta\right) = \frac{\exp\left(z_k\right)}{\sum_{j=1}^{K}\exp\left(z_j\right)}, \quad k=1,...,K.$).
The softmax normalization enforces that the predicted probabilities sum to one and lie in the simplex, enabling a coherent probabilistic interpretation \cite{Goodfellow-et-al-2016}.

The classification model is trained by minimizing the cross-entropy loss \cite{shannon1948} using gradient-based optimizers, such as stochastic gradient descent \cite{bottou2018optimization} or ADAM \cite{kingma2015adam}. 
Adopting the cross-entropy loss is standard in classification tasks because minimizing cross-entropy is equivalent to maximizing the log-likelihood of the training data \cite{hastie2009elements}. 
The cross-entropy loss measures the dissimilarity between the predicted and empirical probability distributions, and is defined as:

\begin{equation}
    \mathcal{L}\left(\theta\right) = -\frac{1}{N} \sum_{i=1}^{N} \log \text{pr} \left( y^{(i)} | \bm{x}^{(i)}; \theta \right).
\end{equation}

In this study, the classification model considered was an MLP with two hidden layers comprising $h=256$ neurons with batch normalization following each ReLU activation function.
Training was conducted using the ADAM optimizer with a learning rate of $10^{-3}$ and weight decay of $10^{-5}$.
The model was set to train for a total of 250 epochs, and early stopping was implemented if the validation loss did not decrease for 40 epochs.
The training dataset comprised 80\% of the complete dataset, and samples were randomly selected through a randomized dataloader with a batch size of 256 samples. 

\subsection{Error Metrics}
\label{ssec:errs}

To assess the performance of the proposed data-driven TSV models, the accuracy (Acc) and F1-score were computed. 
Since the TSV is a discrete quantity assuming a value in $\{-2, -1, 0, +1, +2\}$, PCE predictions are discretized by binning the prediction into its respective discrete value.
The Acc is computed as the number of correct predictions divided by the total number of predictions.
The F1-score is a balance of precision and recall, and is defined as:

\begin{equation}
    F1 = \frac{2\,TP}{2\,TP\,+\,FP\,+\,FN},
\end{equation}
where TP refers to true positives, FP false positives, and FN false negatives.

In addition, the relative root mean squared error (rRMSE) was computed for validating heat stress PCE models, defined as:

\begin{equation}
    \text{rRMSE} = \frac{|| y_{pred} - y_{ref} ||_2}{|| y_{ref} ||_2},
\end{equation}
where $y_{ref}$ represents the reference data and $y_{pred}$ the corresponding model prediction.

\section{Results}
\label{sec:results}

\subsection{Contrasting Variance-Based Sensitivities of Heat Stress and Sensation Metrics}
\label{ssec:SvS}
\begin{figure}[!htbp]
    \centering
    \includegraphics[width=0.99\linewidth]{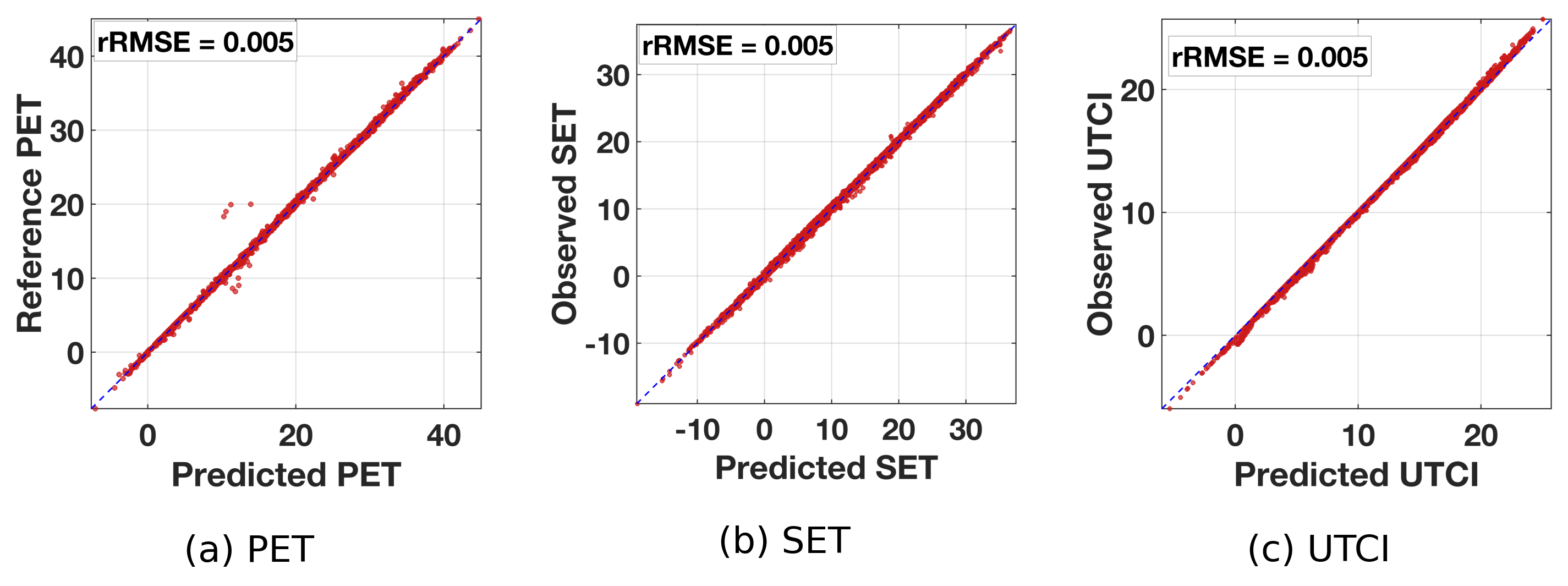}
    \caption{Parity plots comparing fifth-order PC surrogate predictions against reference values for PET, SET, and UTCI (left to right) using the RUROS dataset.}
    \label{fig:heatStress_PCE_makingTheCase}
\end{figure}

We construct PCE models for heat stress (PET, SET and UTCI) and for sensation (TSV) based on the comprehensive RUROS dataset, and calculate the corresponding Sobol sensitivity indices. 
The meteorological input data from RUROS are also used to compute the heat stress metrics in this model evaluation section.
The parity plots for the heat stress metrics, from left to right: PET, SET and UTCI, are presented in Figure \ref{fig:heatStress_PCE_makingTheCase}.
The plots indicate that a $5^{th}$ order ($p=5$) PCE model provides a strong agreement between the predictions and the reference data, with relative root mean squared error values of $0.5\%$.
This is likely attributed to the large sample size and the smooth dependence of the heat stress metrics on the input variables, making the PCE model a high-fidelity surrogate for these heat stress metrics. 
This demonstrates that the PCE representation is sufficiently accurate for variance-based sensitivity analysis of PET, SET, and UTCI. 
We therefore use the resulting Sobol indices as a baseline characterization of how conventional heat stress metrics respond to meteorological forcing, which is then compared with the sensitivity structure of thermal sensation vote models to assess whether subjective thermal sensation is governed by the same dominant drivers, or whether it depends more strongly on higher-order interactions among temperature, humidity, wind, and radiation that are not fully captured by the heat stress metrics.


\begin{figure}[!htbp]
    \centering
    \includegraphics[width=0.9\linewidth]{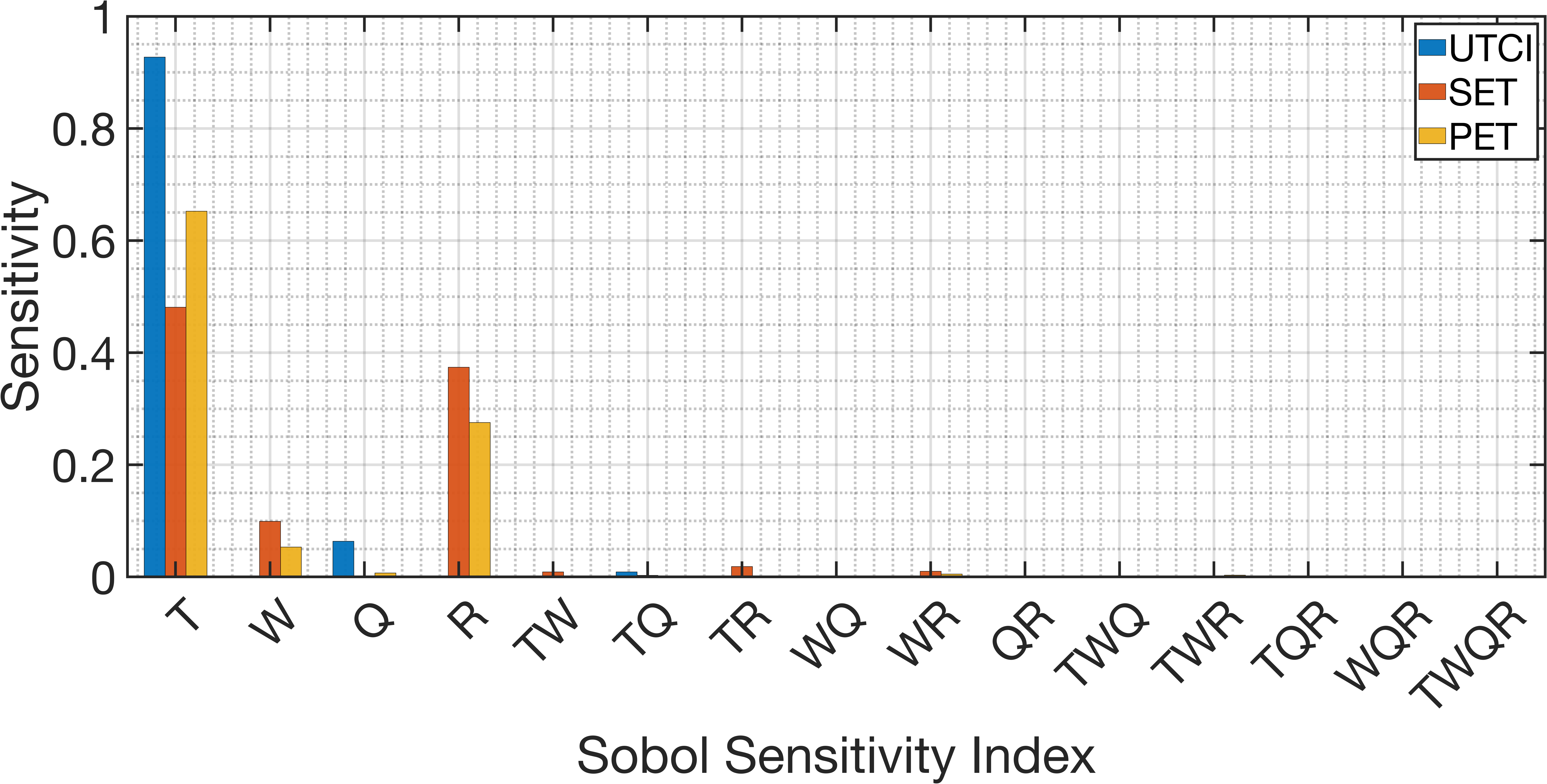}
    \caption{Bar charts illustrating Sobol sensitivity indices computed from the PC surrogates of PET, SET, and UTCI. Here, $T$, $W$, $Q$, and $R$ denote air temperature, wind speed, humidity, and mean radiant temperature, respectively, and concatenated labels denote interaction terms.}
    \label{fig:heatStress_PCE_makingTheCase_sobol}
\end{figure}

Since the PC heat stress models provide high-fidelity heat stress estimates, reliable Sobol sensitivity indices were computed from the PC coefficients and presented in Figure \ref{fig:heatStress_PCE_makingTheCase_sobol}.
The Figure shows the Sobol sensitivity indices for the PCE surrogates of UTCI, SET and PET for different uncertain inputs and combinations thereof as barplots. 
Air and radiant temperatures are the most significant determinants of these indices, though we must underline that the RUROS dataset did not include many high humidity periods under which $Q$ (and $W$ since stronger wind significantly improves evaporative cooling under high humidity conditions), so these sensitivities are specific to the parameter space of RUROS.
The bar charts show that heat stress variability is primarily dependent on individual contributions of the variable inputs, and cross-interaction terms have a negligible effect on heat stress. 
This raises the question of whether perceived heat exhibits a similar sensitivity to meteorological variables as heat stress? 

\begin{figure}[t]
    \centering
    \includegraphics[width=\linewidth]{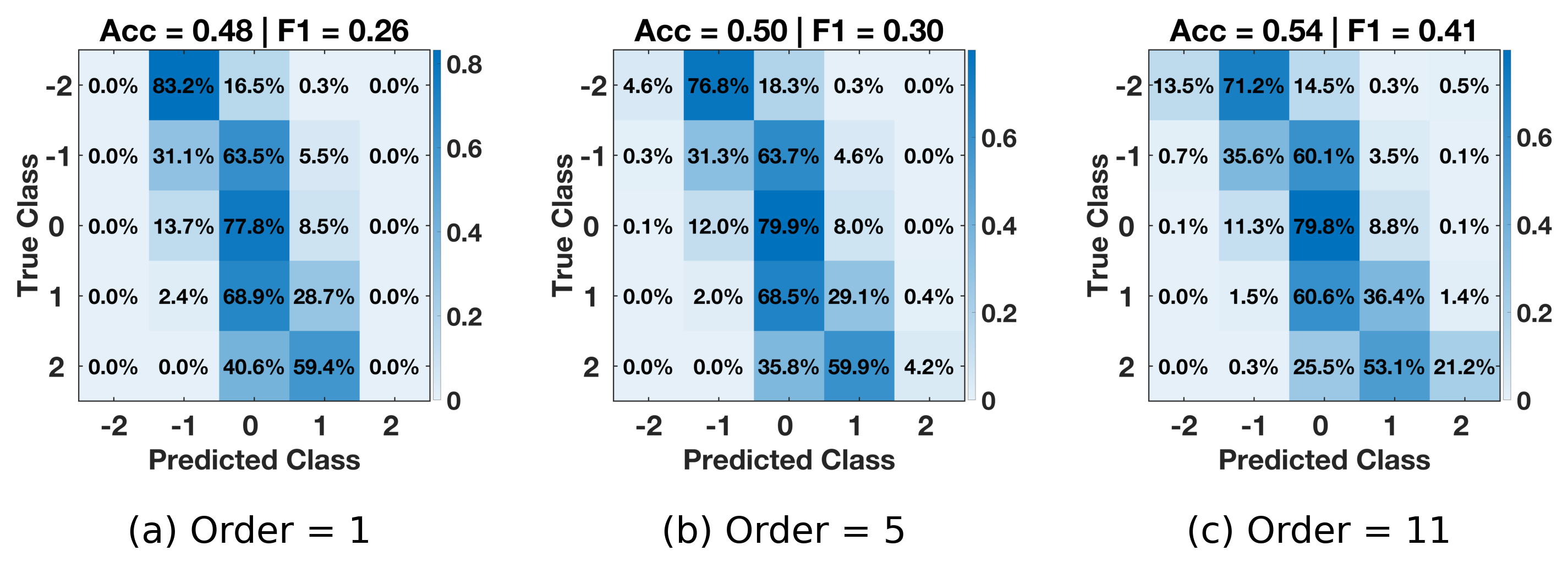}
    \caption{Confusion matrices for TSV classes predicted by polynomial chaos surrogates of increasing total order: order 1 (left), order 5 (center), and order 11 (right). Continuous PC predictions were binned into the five TSV classes $\{-2,-1,0,1,2\}$.}
    \label{fig:heatSens_PCE_makingTheCase}
\end{figure}

To address this question, we construct PC-based surrogates for TSV using the same dataset, and compute the Sobol sensitivities.
Note that we acknowledge that TSV provides discrete sensation categories and PCE models are best suited for continuous response functions \cite{Canuto2006}; however, in agreement with past research \cite{GOLASI2018, FENG2021100938, POTCHTER2018390} and for the sake of estimating Sobol sensitivities, this can be relaxed provided that overfitting is avoided.  
To this end, we rely on LASSO regression to regularize the regression problem and ensure that the number of coefficients of the PCE model is well below the number of data samples used for training. 
For classification-based evaluation, we post-process the continuous PCE predictions by discretizing them into five ordinal classes using fixed threshold intervals corresponding to the labels $\{-2,-1,0,+1,+2\}$.
Figure \ref{fig:heatSens_PCE_makingTheCase} presents confusion matrices for PCE models of order 1, 5 and 11, respectively where each cell reports the percentage of the model predictions ($x$ axis value) given the total number of true values ($y$ axis). The sum of each row is 100\% and a perfect model would only have entries of 100\% along its main diagonal (upper left to lower right).
The plots indicate that higher-order models enhance prediction reliability, as shown by the increased accuracy and F1 score, which is expected.
Note that, in order to avoid overfitting, two strategies were adopted: first, the number of coefficients of the PCE was kept smaller than the number of data samples by specifying a suitable total order, and second, LASSO employs $\ell_1$ regularization, which promotes sparse PCE models that are less susceptible to overfitting.

\begin{figure}[!htbp]
    \centering
    \includegraphics[width=0.85\linewidth]{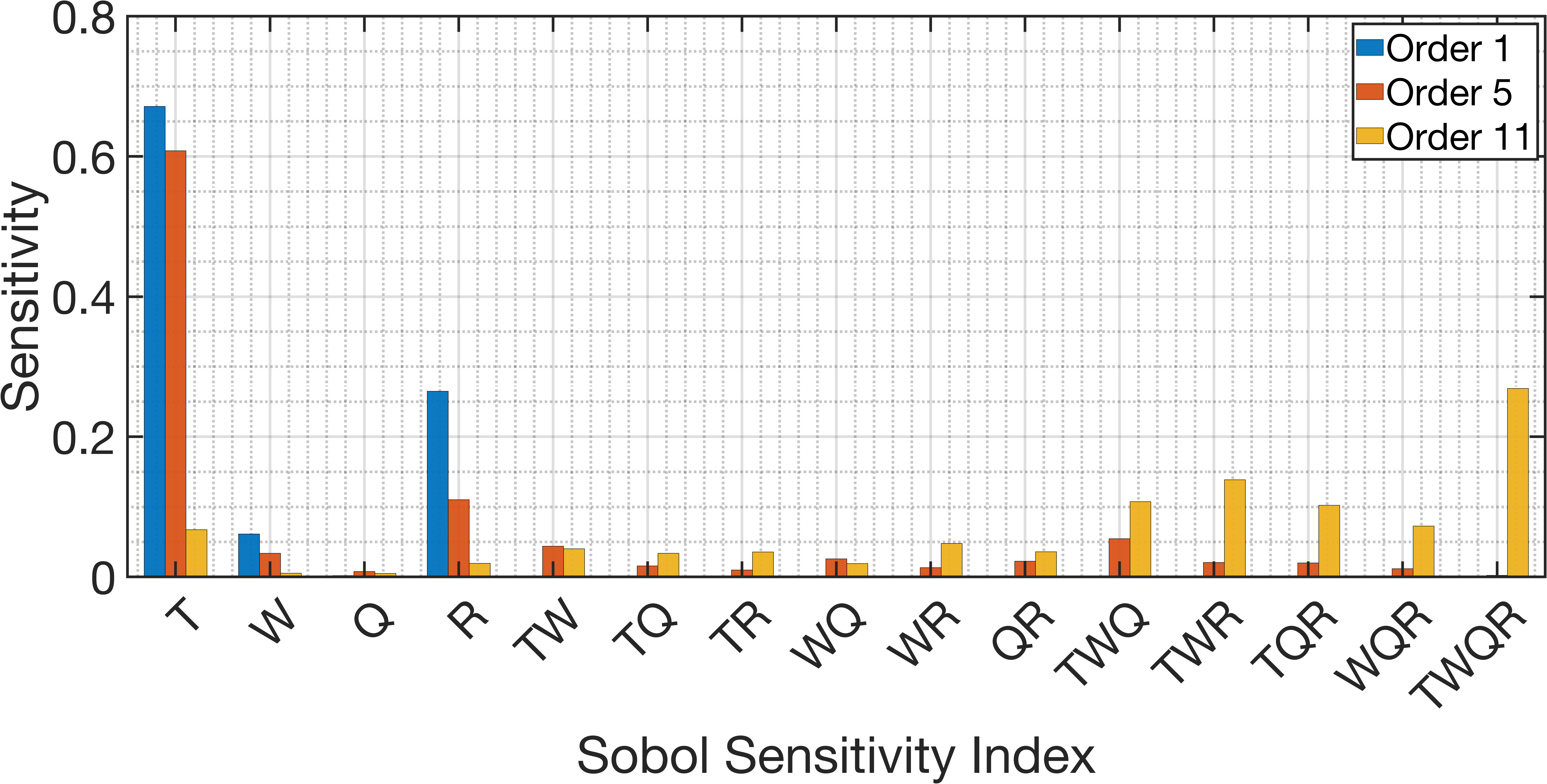}
    \caption{Sobol sensitivity indices for TSV obtained from polynomial chaos surrogates of order 1, 5, and 11.}
    \label{fig:heatSens_PCE_makingTheCase_sobol}
\end{figure}

Sobol sensitivities are then computed using the heat sensation PCE models, and barplots for these sensitivities for PCE models of order 1, 5 and 11 are presented in Figure \ref{fig:heatSens_PCE_makingTheCase_sobol}.
Note that these sensitivity indices converge to a limiting behavior as the PC order increases; however, the PC total order should be maintained such that the number of coefficients is smaller than the number of data samples to take precautions to avoid overfitting.
The plots indicate that a first-order model cannot capture cross-interaction terms (by construction), and higher-order interactions appear with higher PC orders. 
For the 11$^{\text{th}}$-order model (most reliable predictions), TSV exhibits dominant sensitivities to higher-order interaction terms, while several singular effects become comparatively small.
This structure shows a sharp contrast with stress-based metrics, which tend to emphasize dominant single-variable contributions. 
In the TSV model, the influence of humidity (Q), for instance, appears weak as a standalone driver but emerges strongly through interaction terms, consistent with the physical mechanism that evaporative cooling depends jointly on temperature gradients and wind-driven convective transport.
These results suggest that thermal sensation is governed primarily by coupled meteorological processes rather than isolated environmental drivers. 

We note that both the heat stress and sensation sensitivities were calculated in the same input parameter space, so the differences are inherently linked to the simple input-output maps in the stress models that are in contrast to the more complex and intertwined maps for thermal sensation. This contrast can only increase (or remain the same) if the input parameter space is broadened.

\subsection{Heat Sensation: Estimation and Classification}
\label{ssec:TSV_Global2}

\begin{figure}[!htbp]
    \centering
    \includegraphics[width=0.95\linewidth]{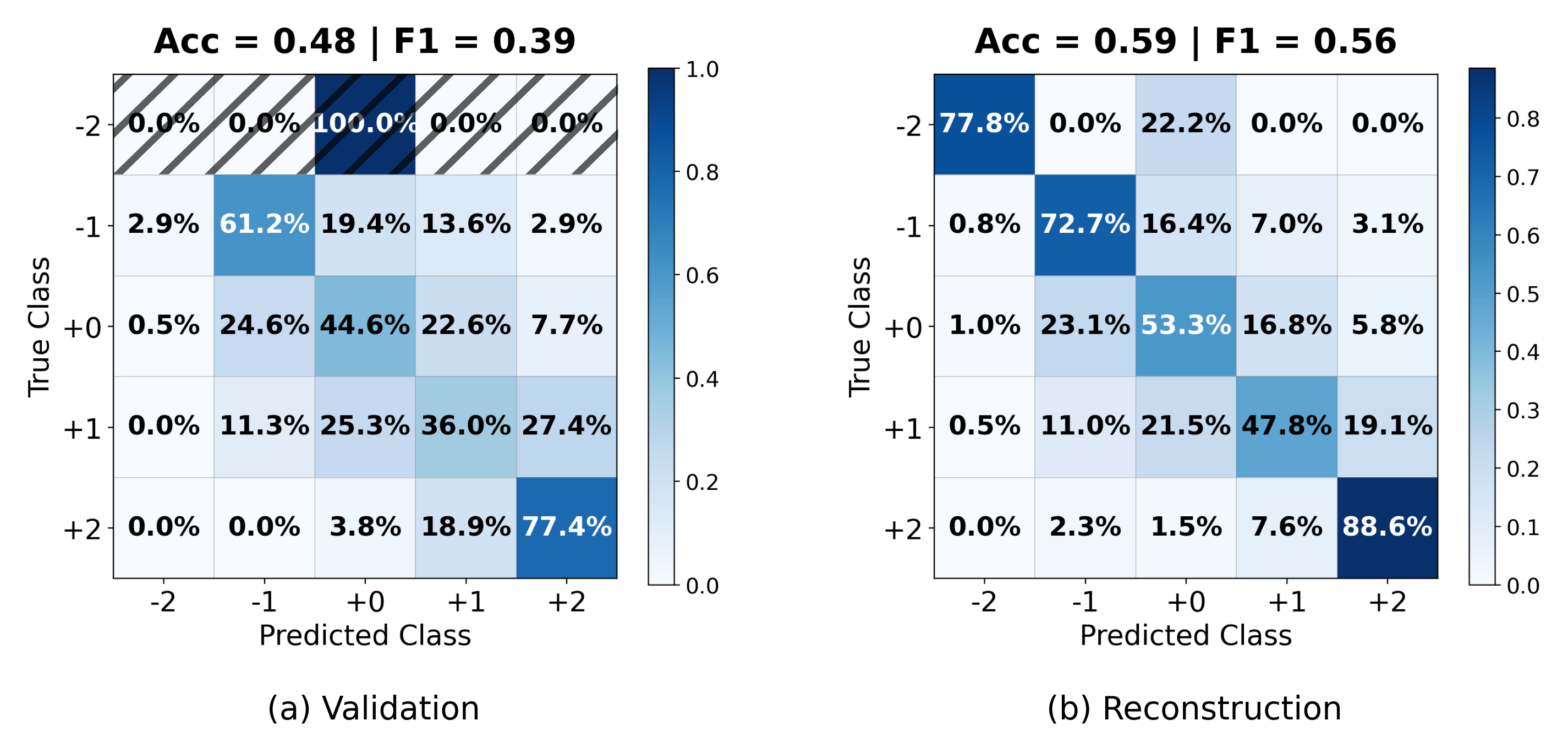}
    \caption{Confusion matrices for the neural network classification model trained on the RUROS dataset and evaluated on the validation subset (left) and the full dataset (right). The hatched row represents the case with less than 10 data samples, making the results statistically insignificant.}
    \label{fig:logReg_cm}
\end{figure}

Recognizing the limitations of continuous models in fitting categorical datasets, we also explore classification as an alternative solution for building heat sensation models. 
The combined RUROS dataset was used to train a classification model that maps meteorological conditions to TSV classes via an MLP. 
Figure \ref{fig:logReg_cm} shows confusion matrices for the MLP's predictions for the validation (left, 1,806 samples) and entire (right, 9,028 samples) datasets.
The plots indicate that the model provides reliable predictions that are close to the reference, with errors mostly resulting in predictions in the adjacent class.
The Figure shows that the MLP predictions achieve an accuracy of 48$\%$ on the validation dataset and $59\%$ on the comprehensive dataset, and respective F1-scores of 0.39 and 0.56, which is comparable with the best PCE model.
In contrast to the PCE model, which offers continuous and unbounded predictions, the MLP predictions are probabilities for the ordinal TSV values, allowing the MLP model to inherit subjectivity elements of TSV. 
In particular, since the TSV data varies from one person's perception of thermal comfort, it is idiosyncratic.
The MLP estimates the probability of the TSV classes, hence expressing the subjective variability among individuals in the dataset; the class with the highest probability is then considered as the most likely prediction (i.e., what a majority of individuals would report).

\begin{figure}[!htbp]
    \centering
    \includegraphics[width=0.45\linewidth]{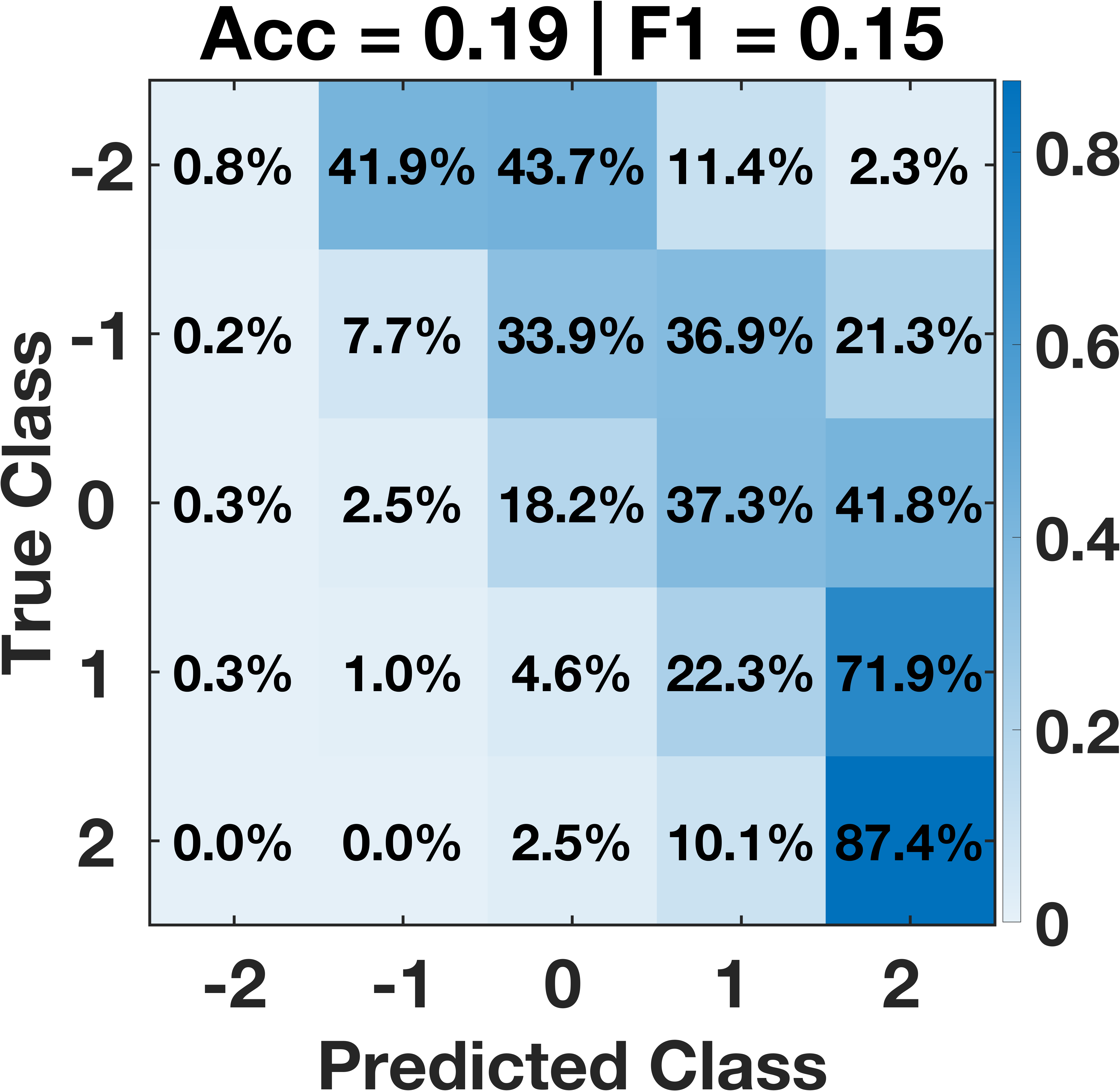}
    \caption{Confusion matrix for TSV classes predicted by the GOCI-based linear model after binning its continuous output into discrete TSV classes.}
    \label{fig:GOCI_ConfMat}
\end{figure}

Further illustrating the advantage of the MLP model, the GOCI  predictions of TSV were computed for the same dataset. 
The GOCI predictions were also binned into discrete intervals, and the confusion matrix is illustrated in Figure \ref{fig:GOCI_ConfMat}. 
The Figure shows that GOCI predictions overestimate TSV predictions by 1 class in the warmer direction, which is indicated by the larger number of off-diagonal predictions, therefore, it might be too conservative, overpredicting thermal discomfort. 
The confusion matrix indicates that the accuracy of the GOCI model on the validation dataset is around $19\%$ (compared to $54\%$ with the 11$^{th}$ order PCE model) and achieves an F1-score of $0.15$ (compared to $0.41$ with the 11$^{th}$ order PCE model).

Because the MLP classifier returns class probabilities, it can be applied directly to gridded meteorological fields to estimate the most likely thermal sensation response at large scales.
Moreover, since the classification MLP model provides reliable predictions on the validation dataset, it can then be applied on a global scale to predict how an average European (since the RUROS surveys are all done in Europe) would perceive heat. 
We collected T, Q, and W from ERA-5 dataset \cite{Hersbach2020} and R from ERA-Heat dataset (Di Napoli et al., 2020), for August 12$^{th}$, 2024, which represents typical meteorological conditions of a heat wave period (August 11$^{th}$-13$^{th}$, 2024) where many European countries (e.g., U.K. \cite{bbc_uk_weather_2024}, Belgium \cite{meteo_be_augustus_2024}, France \cite{extremeweatherwatch_montdemarsan_2024}, Germany \cite{wetterzentrale_de_obs_20240813}) recorded the highest air temperature during the 2024 summer heatwave. 
Meteorological conditions for 2 pm local time, daytime average, and 24-hour average were then extracted. 
For the sake of brevity, only TSV predictions corresponding to the 2 pm local conditions are presented in this study, where results for the daytime and 24-hour averages showed similar TSV patterns but are closer to TSV=0 since the nighttime is cooler and dark (low radiant temperature).
The meteorological inputs were propagated through the MLP that outputs the logits (or probability classes) from which the class with the highest probability is the model's TSV prediction.

\begin{figure}
    \centering
    \includegraphics[width=0.99\linewidth]{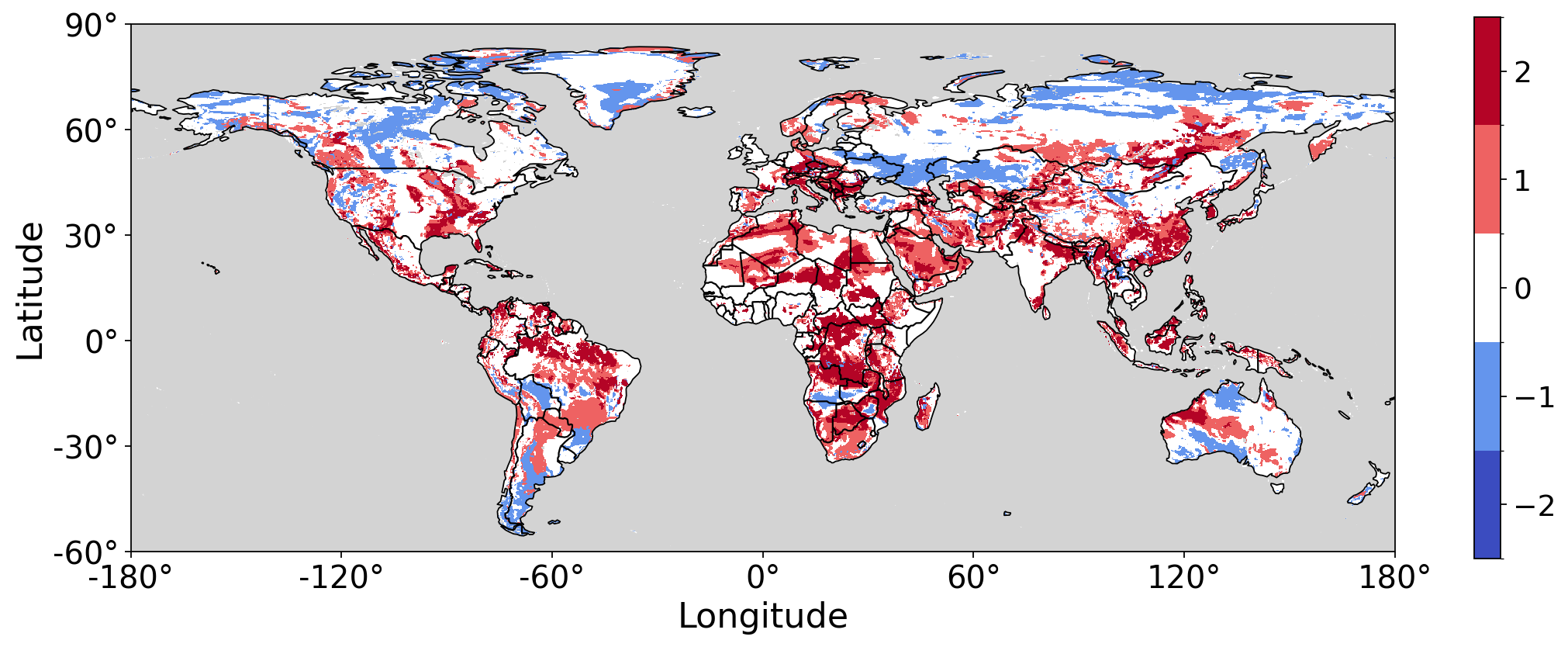}
    \caption{Global map of the most likely TSV class predicted by the neural network classification model from the meteorological conditions on 12 August 2024 at 2pm local time.}
    \label{fig:gloablTSVpred}
\end{figure}

Figure \ref{fig:gloablTSVpred} illustrates the MLP's most likely TSV prediction. 
Warm to very warm sensation (TSV = +1 and +2) dominates most of the tropics, subtropic and continental mid-latitudes of the northern hemisphere.
The highest-probability +2 class is concentrated over the hottest land regions, including North Africa and the Arabian Peninsula, large parts of southern Africa, portions of tropical South America, interior North America, and several regions across South and East Asia (note that this is for the weather of the particular day we model here, with the data representative of the local 2 pm conditions).
Furthermore, the thermal discomfort experienced in various European cities due to the heat wave was captured by the model, where Italy, South and East Germany, and Greece, among others, are predicted to have very warm (TSV=+2) conditions, whereas warm conditions (TSV=+1) are predicted for Spain.  
In contrast, negative TSV classes are mainly confined to high-latitude and winter-hemisphere regions, including northern Canada, Greenland, northern Eurasia, and southern South America.
Neutral conditions appear primarily in transitional regions between these hot and cold regimes, indicating that mid-day thermal neutrality is spatially limited relative to warm sensations on this date.


\begin{figure}
    \centering
    \includegraphics[width=0.80\linewidth]{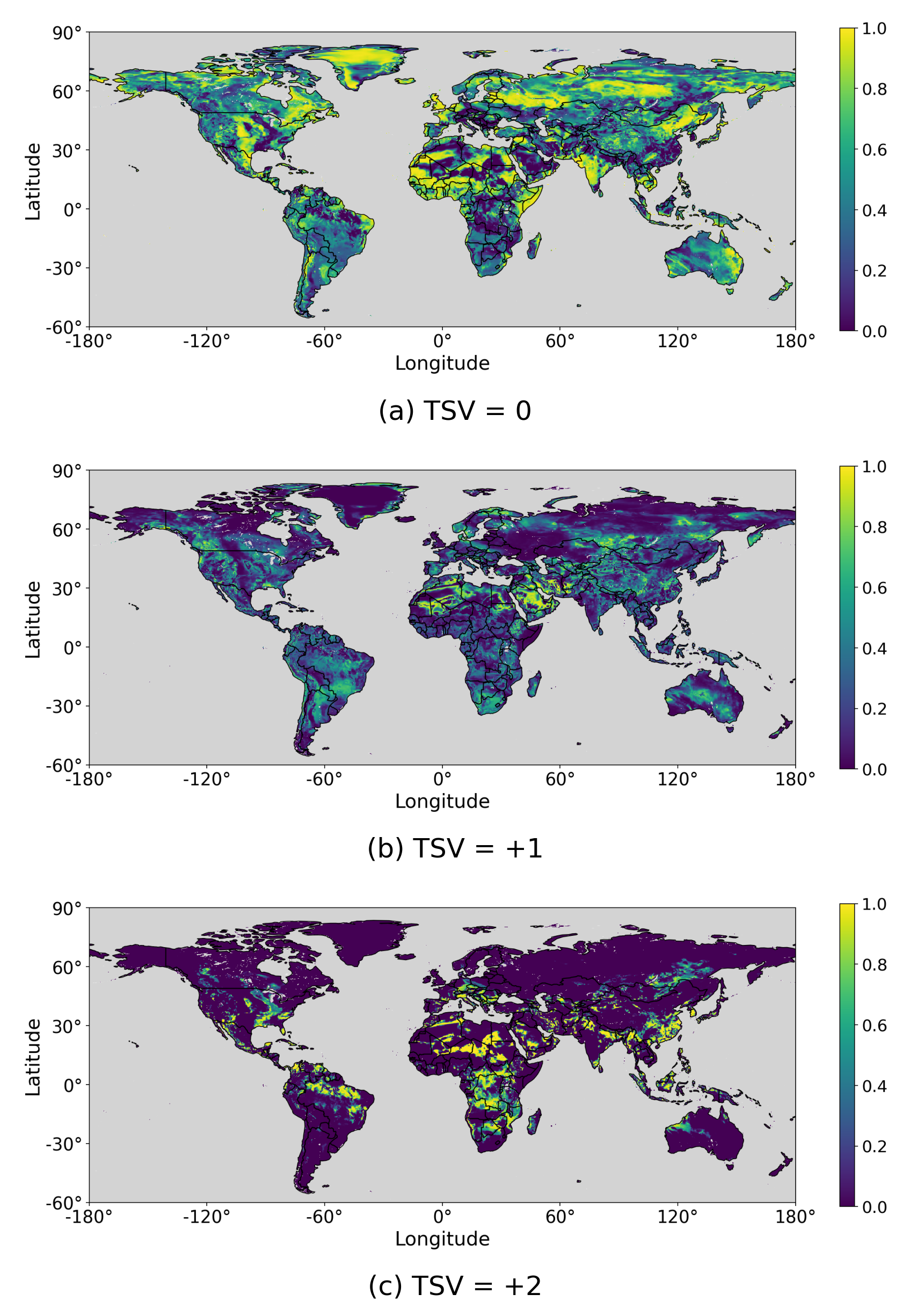}
    \caption{Global probability maps from the MLP-based classification model for (a) TSV $=0$, (b) TSV $=+1$ and (c) TSV $=+2$ on 12 August 2024. These logits-derived probabilities provide a probabilistic representation of subjective heat perception, identifying where neutral conditions are most likely and where very hot sensation is most probable.}
    \label{fig:logits}
\end{figure}

Furthermore, a more detailed view of the global structure can be provided through the logit maps, which represent the class probabilities output by the neural network. 
For illustration, Figure \ref{fig:logits} shows the probability of perceived heat sensation falling under classes 0, +1 or, +2. 
These maps not only capture the argmax class (most likely prediction), but also the spatial gradients in the classification confidence, and are particularly useful near class boundaries, where multiple TSV classes remain plausible and the most likely TSV class changes rapidly in the parameter space.
The probability of TSV = 0 is spatially broad and remains appreciable across many transition zones, highlighting overlap between neighboring sensation classes. 
The TSV = +1 field forms wide belts around the hottest regions and is most prominent where conditions are warm but not at the far end of the model response range. 
On the other hand, TSV = +2 probabilities are more localized, with compact maxima over the hottest continental and tropical regions identified in Figure \ref{fig:gloablTSVpred}. 

\subsection{City-Level Heat Sensation} 
\label{ssec:TSV_Regional}

\begin{figure}
    \centering\includegraphics[width=0.98\linewidth]{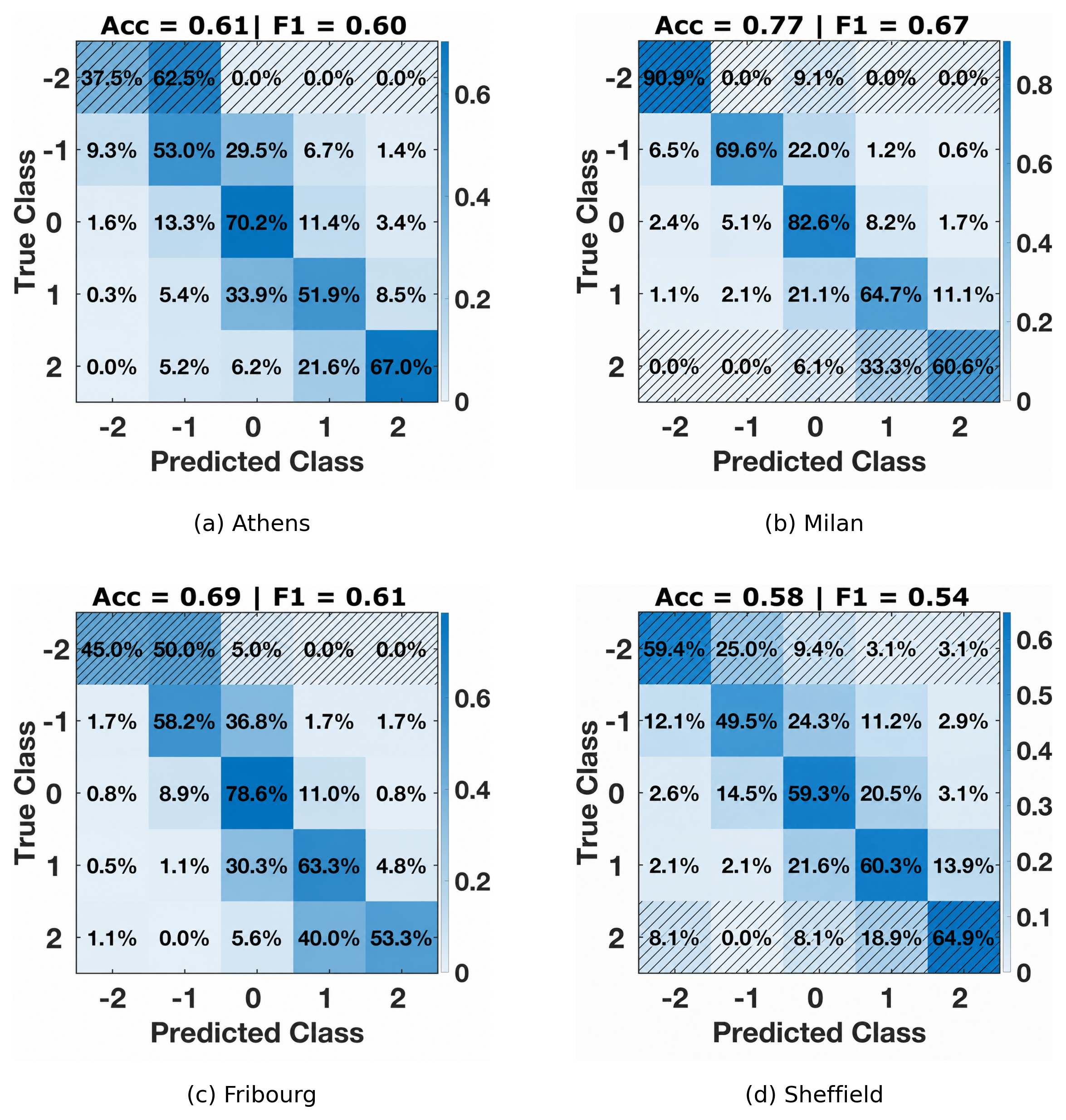}
    \caption{Confusion matrices for TSV classes predicted by ninth-order PCE surrogates for (a) Athens, (b) Milan, (c) Fribourg, and (d) Sheffield. The PCE predictions were binned into the five ordinal TSV classes $\{$-2, -1, 0, +1, +2$\}$. The shaded rows represent cases with less than 10 data samples, meaning they are not statistically representative.}
    \label{fig:regionalPCE_cm}
\end{figure}

We then explore the spatial heterogeneity in perceived heat. 
We construct city-specific PCE surrogates for four cities covered in the RUROS database: Athens, Milan, Fribourg and Sheffield. 
Here, we restrict the analysis to 9$^{th}$ order PCE surrogates to avoid the potential for overfitting because the size of the subdatasets for individual cities is much smaller than that of the comprehensive RUROS dataset. 
Figure \ref{fig:regionalPCE_cm} presents confusion matrices for the binned PCE TSV predictions against the observations for the cities mentioned above. 
The figure shows that these city-level PCE surrogates reproduce the observed TSV classes with very good skill, with accuracies and F1-scores of 0.61/0.6 for Athens, 0.77/0.67 for Milan, 0.69/0.61 for Fribourg, and 0.58/0.54 for Sheffield.
In all four cities, the largest values remain on or near the diagonal, and most errors occur between nearby TSV categories, indicating that the PC surrogates are reasonable surrogates for thermal sensation.

\begin{figure}
    \centering
    \includegraphics[width=0.95\linewidth]{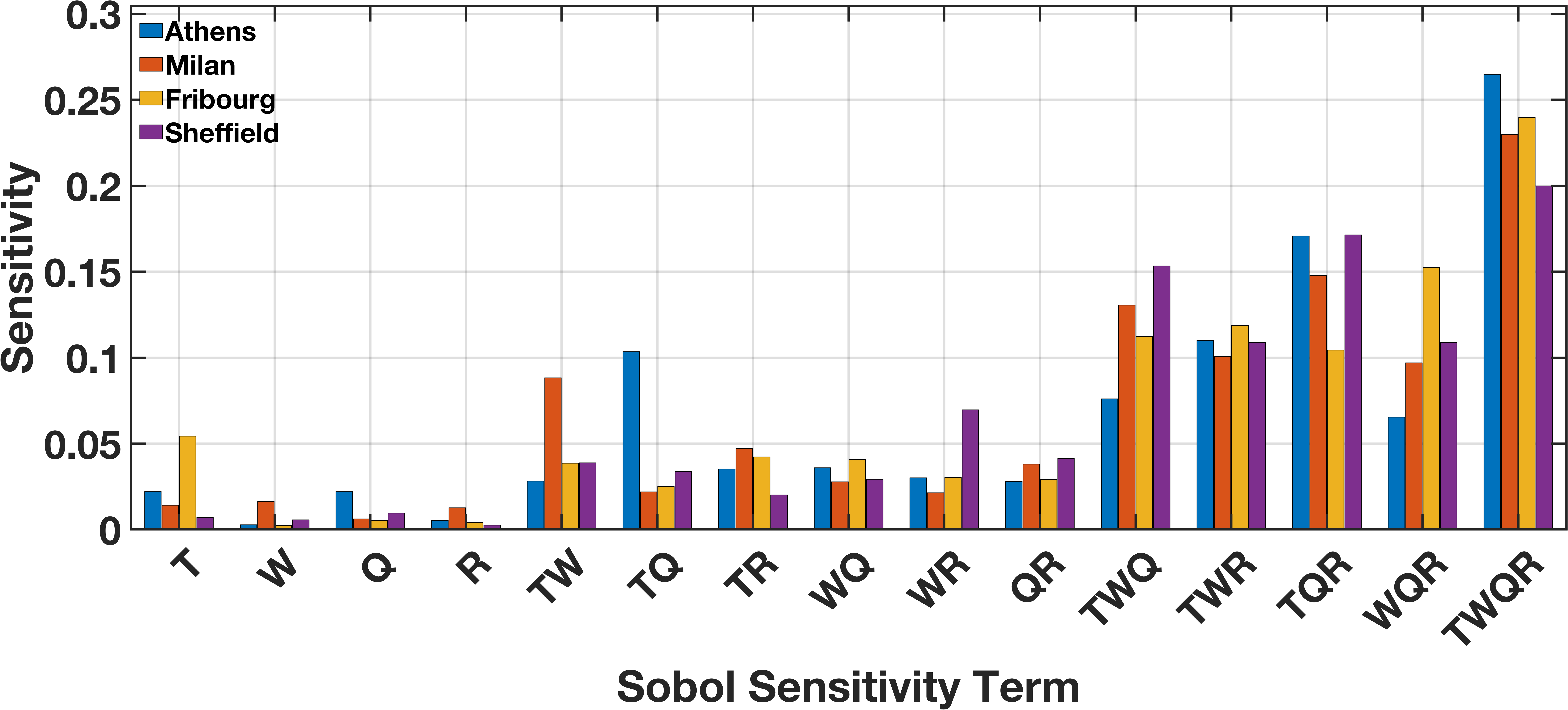}
    \caption{Sobol sensitivity indices for TSV obtained from the ninth-order city-specific PCE surrogates for Athens, Milan, Fribourg, and Sheffield, as indicated by the legend.}
    \label{fig:regionalPCE_sens}
\end{figure}

Figure \ref{fig:regionalPCE_sens} illustrates bar charts of the Sobol sensitivity indices corresponding to thermal sensation at the city-level. 
The plot confirms the previous conclusion from the entire RUROS dataset: individual city-scale thermal sensation is also controlled predominantly by interactions among T, Q, W, and R, rather than by isolated first-order effects. This implies that a temperature-dominant narrative of urban climate would ignore important interactions and impacts from other meteorological variables on human thermal sensation. 
Across all four cities, the four-way interaction term (TWQR) is the single largest contributor, or among the largest contributors, to the total variance.
Three-way terms also account for a substantial share of the variance, while the individual effects of T, Q, W, and R remain comparatively small. The non-linearity of the 'best model' determined through PCE encodes the interaction of these meteorological variables in shaping thermal comfort.
At the same time, the relative importance of specific combinations varies across cities.
For instance, Athens shows a comparatively stronger TQ contribution, Milan a stronger TW contribution, Fribourg a somewhat larger first-order temperature effect together with strong WQR, and Sheffield elevated WR, TWQ, and TQR contributions. 
These differences indicate that the meteorological controls on thermal sensation are not spatially universal, but instead depend on local geoclimatic conditioning and on how environmental variables combine in a given setting.

\begin{figure}
    \centering
    \includegraphics[width=0.95\linewidth]{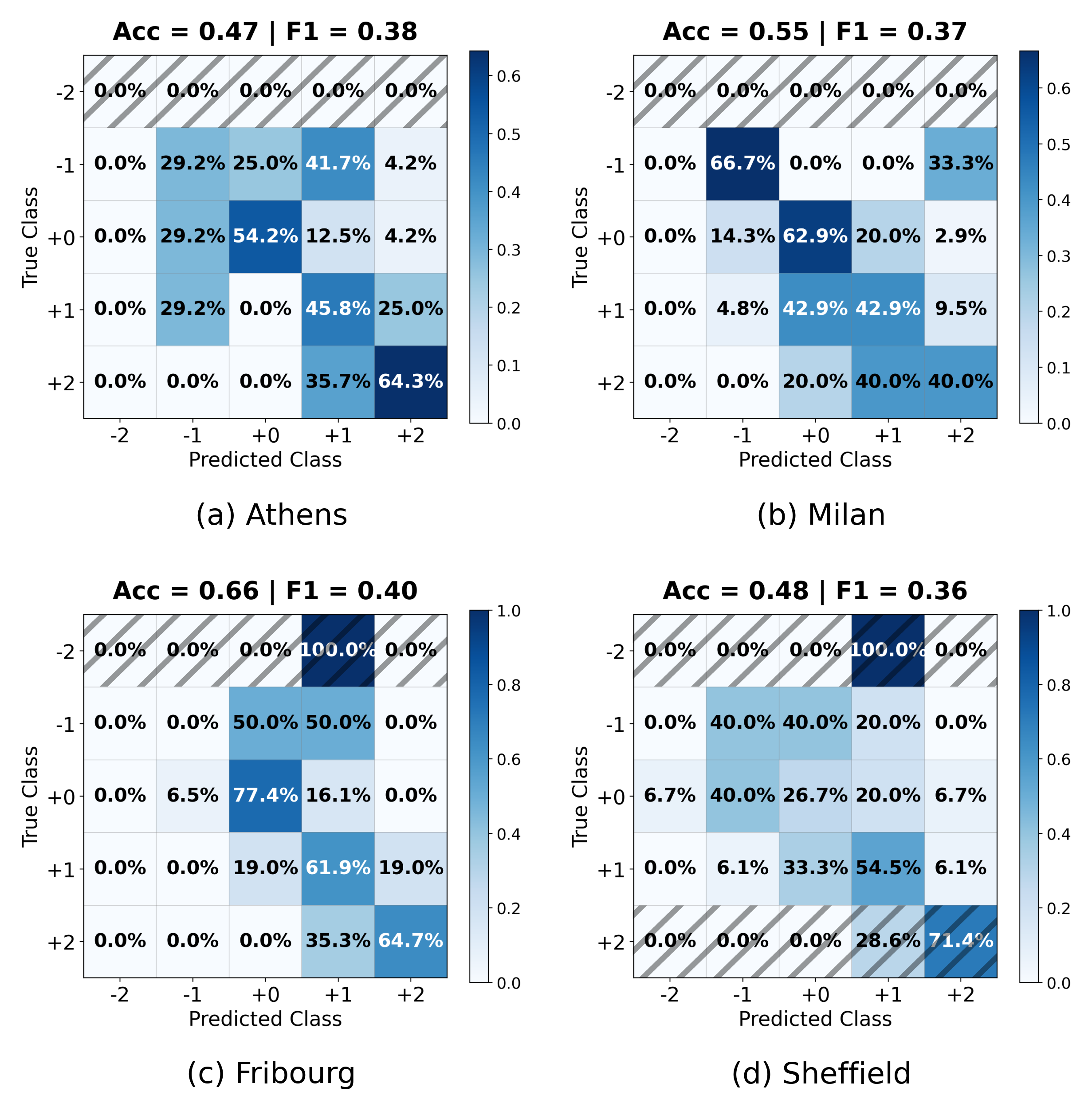}
    \caption{Confusion matrices for TSV classes predicted by city-specific MLP-based classification models for (a) Athens, (b) Milan, (c) Fribourg, and (d) Sheffield. The shaded rows represent cases with less than 10 data samples, meaning they are not statistically representative.}
    \label{fig:regionalMLP_cm}
\end{figure}

Similarly, MLP-based classifiers can be trained at the city-level. 
Figure \ref{fig:regionalMLP_cm} represents confusion matrices for the MLP classification predictions for Athens, Milan, Fribourg and Sheffield.
The city-specific MLP classifiers show the same behavior in terms of local specificity, but with slightly less reliable predictive skill than the PCE surrogates. 
The corresponding accuracies/F1 scores are 0.47/0.38 for Athens, 0.55/0.37 for Milan, 0.66/0.40 for Fribourg, and 0.48/0.36 for Sheffield. 
While these models are less reliable than the city-specific PCE surrogates (lower F1 score, meaning more false positives and negatives) because of the availability of fewer data samples, a substantial fraction of predictions still falls on or close to the diagonal, with the neutral and warm classes generally captured more reliably than the sparsest tail end categories. 
The noisier class-wise behavior corresponds to smaller sample sizes and stronger class imbalance at the city scale. 
Overall, the results for the PCE and MLP surrogates suggest that TSV surrogates are achievable at the city scale, but predictive skill and sensitivity structure vary consequentially across cities. 
This suggests that human thermal perception is shaped by locally-conditioned combinations of meteorological drivers and population adaptation, rather than by a universal rule.

\clearpage

\section{Discussion and Conclusions}
\label{sec:conc}

We developed a data-driven framework for modeling human thermal sensation that explicitly distinguishes heat perception from traditional heat-stress metrics.
Using thermal sensation votes survey data, we showed that variance-based sensitivities of common heat-stress indices differ fundamentally from those governing reported thermal sensation.
This mismatch demonstrates that stress-based indices, while physiologically meaningful, might not be appropriate surrogates for human heat perception.
This is a consequential finding since health outcomes and their mitigation are better quantified using stress metrics \cite{URBAN2021111227, Guo_2024}, while comfort goals (more livable cities, reduced indoor cooling demand) are best reflected through comfort metrics. 

To address this disagreement, we formulated thermal sensation prediction using two complementary surrogate modeling approaches.
Polynomial chaos expansions (PCEs) were employed to represent thermal sensation as a function of environmental drivers, enabling direct Sobol sensitivity analysis and explicit identification of influential variables and their higher-order interactions.
This formulation yields a surrogate that reveals which inputs, and combinations thereof, govern perceived heat.
In parallel, multilayer perceptron (MLP) classifiers were trained to capture the nonlinear and subjective nature of thermal sensation, providing a flexible predictive model unconstrained by linearity.

\textit{A priori} analysis using the PCE framework reveals that the dominant sensitivities of thermal sensation differ markedly from those of heat-stress metrics.
Interaction terms that dominate heat perception are often secondary or negligible for stress-indices, with perception-based models emphasizing combinations of temperature, radiation, humidity, and wind.
These sensitivity structures provide direct insight into why heat-stress indices do not directly correspond to sensation, and offer a rigorous and reproducible pathway for constructing new, perception-oriented thermal models informed by data.

In a more applied setting, the MLP models demonstrate strong predictive skill across both regional and global datasets, indicating their suitability for capturing the subjective and nonlinear nature of thermal sensation.
Across climates, the models remain robust while naturally adapting to local environmental regimes.
Contrasting PCE-based Sobol sensitivity indices across different cities shows that no single environmental driver universally governs thermal sensation, reinforcing the limitations of globally uniform indices when used to infer human experience.

The combined use of interpretable PCE surrogates and neural network models establishes a new modeling paradigm for human thermal perception.
Rather than fitting sensation as a byproduct of heat stress, this approach treats perception as a primary quantity, with its own governing structure and uncertainties.
This distinction is critical for applications where human experience is the quantity of interest, including climate impact assessment, indoor and outdoor thermal comfort analysis, and human-centric climate services.
Further granularity can also be attained with these neural TSV models: since they capture the subjective nature of heat sensation, these models can be deployed onto wearable electronics and smart devices, allowing them to be trained by and tailored to specific individuals.
Finally, the authors intend to expand on the RUROS dataset to collect heat sensation data from various locations around the world to better understand the drivers of heat sensation across different age groups, body types, and geoclimates.  
Furthermore, by generating an updated dataset, one can investigate whether we are getting better adapted to the warming climate, and whether the drivers of heat sensation have changed over time.

%
%

\section{Data availability}
The Rediscovering the Urban Realm and Open Spaces (RUROS) survey database is openly available through Zenodo (Nikolopoulou \& Lykoudis, 2004; repository version deposited in 2024; DOI: \href{https://doi.org/10.5281/zenodo.14275070}{\nolinkurl{10.5281/zenodo.14275070}}). 
The complete archive contains 9,271 survey records. 
ERA5 hourly single-level meteorological fields were obtained from the Copernicus Climate Data Store (DOI: \href{https://doi.org/10.24381/cds.adbb2d47}{\nolinkurl{10.24381/cds.adbb2d47}}), and mean radiant temperature was obtained from the ERA5-HEAT dataset (DOI: \href{https://doi.org/10.24381/cds.553b7518}{\nolinkurl{10.24381/cds.553b7518}}). 
The relevant software references and DOIs are: pythermalcomfort: \href{https://doi.org/10.1016/j.softx.2020.100578}{\nolinkurl{10.1016/j.softx.2020.100578}} and thermofeel: \href{https://doi.org/10.1016/j.softx.2022.101005}{\nolinkurl{10.1016/j.softx.2022.101005}}.

\section{Competing interest}
The authors declare there are no conflicts of interest for this manuscript.

\section{Funding}
This research has been supported by the National Oceanic and Atmospheric Administration (US Department of Commerce grant no. NA23OAR4320198) and Princeton University through the Cooperative Institute for Modeling the Earth System.
A.H. is supported by the Gordon and Betty Moore Foundation’s Postdoctoral Fellowship.
E.B.Z. and X.H. are supported by M. S. Chadha Center for Global India at Princeton University. 
X.H. is supported by the High Meadows Environmental Institute at Princeton University through the generous support of the William Clay Ford, Jr. ‘79 and Lisa Vanderzee Ford ‘82 Graduate Fellowship Fund.  

\bibliographystyle{unsrt}
\bibliography{reference}

\end{document}